\documentclass[aps,prc,preprint,superscriptaddress,showpacs,preprintnumbers]{revtex4-1}
\usepackage{graphicx}
\usepackage{dcolumn}
\usepackage{bm}
\usepackage{longtable}
\usepackage[colorlinks]{hyperref}
\begin{document}
{\setlength{\tabcolsep}{6pt}
\title{Measurement of relative isotopic yield distribution of even-even fission fragments
from $^{235}$U($n_{th}$,$f$) following $\gamma$ ray spectroscopy}
\author{Aniruddha Dey}
\email[]{deyaniruddha07@gmail.com}
\address{Nuclear Physics Division, Bhabha Atomic Research Centre, Mumbai 400 085, India}
\address{Department of Physics, Siksha Bhavana, Visva-Bharati University, Santiniketan, West Bengal 731 235, India}
\author{D.C. Biswas}
\email[]{dcbiswas11@gmail.com}
\address{Nuclear Physics Division, Bhabha Atomic Research Centre, Mumbai 400 085, India}
\author{A.~Chakraborty}
\address{Department of Physics, Siksha Bhavana, Visva-Bharati University, Santiniketan, West Bengal 731 235, India}
\author{S.~Mukhopadhyay}
\address{Nuclear Physics Division, Bhabha Atomic Research Centre, Mumbai 400 085, India}
\author{A.~K.~Mondal}
\address{Department of Physics, Siksha Bhavana, Visva-Bharati University, Santiniketan, West Bengal 731 235, India}
\author{L.~S.~Danu}
\address{Nuclear Physics Division, Bhabha Atomic Research Centre, Mumbai 400 085, India}
\author{B.~Mukherjee}
\address{Department of Physics, Siksha Bhavana, Visva-Bharati University, Santiniketan, West Bengal 731 235, India}
\author{S.~Garg}
\address{School of Physics and Astronomy, Shanghai Jiao Tong University, Shanghai 200240, P. R. China}
\author{B.~Maheshwari}
\address{Amity Institute of Nuclear Science and Technology, Amity University UP, Noida 201 313, India}
\author{A.~K.~Jain}
\address{Amity Institute of Nuclear Science and Technology, Amity University UP, Noida 201 313, India}
\author{A.~Blanc}
\address{Institut Laue-Langevin, 71 Avenue des Martyrs, 38042 Grenoble CEDEX 9, France}
\author{G.~de~France}
\address{GANIL, BP 55027, F-14076 Caen Cedex 5, France}
\author{M.~Jentschel}
\address{Institut Laue-Langevin, 71 Avenue des Martyrs, 38042 Grenoble CEDEX 9, France}
\author{U.~K$\ddot{o}$ster}
\address{Institut Laue-Langevin, 71 Avenue des Martyrs, 38042 Grenoble CEDEX 9, France}
\author{S.~Leoni}
\address{Universit$\grave{a}$ degli Studi di Milano, I-20133 Milano, Italy}
\author{P.~Mutti}
\address{Institut Laue-Langevin, 71 Avenue des Martyrs, 38042 Grenoble CEDEX 9, France}
\author{G.~Simpson}
\address{LPSC, 53 Avenue des Martyrs, 38026 Grenoble, France}
\author{T.~Soldner}
\address{Institut Laue-Langevin, 71 Avenue des Martyrs, 38042 Grenoble CEDEX 9, France}
\author{C.~A.~Ur}
\address{INFN Sezione di Padova, I-35131 Padova, Italy}
\author{W.~Urban}
\address{Faculty of Physics, University of Warsaw, PL 02-093 Warszawa, Poland}

\date{\today}

\begin{abstract}

A detailed investigation on the relative isotopic distributions has been carried out for the first time in case of 
even-even correlated fission fragments for the $^{235}$U($n_{th}$,$f$)
fission reaction. High-statistics data were obtained in a prompt $\gamma$ ray spectroscopy
measurement during the EXILL campaign at ILL, Grenoble, France.
The extensive off-line analysis of the coincidence data have been carried out
using four different coincidence methods.
Combining the results from 2-dimensional $\gamma-\gamma$ and 
3-dimensional $\gamma-\gamma-\gamma$ coincidence analysis, 
a comprehensive picture of the relative isotopic yield distributions
of the even-even neutron-rich fission fragments has emerged.
The experimentally observed results have been substantiated by the theoretical calculations 
based on a novel approach of isospin conservation, and a reasonable agreement has been obtained.
The calculations following the semi-empirical GEF model have also been carried out. The results from the GEF
model calculations are found to be in fair agreement with the experimental results.
\end{abstract}

\maketitle

\section{Introduction}

Nuclear fission is an extremely complex phenomenon with 
the collective involvement and rearrangement of the participating nucleons
in a multi-dimensional deformation space. In general, a nucleus splits 
into two fragments of comparable sizes during a fission process. 
The process is also involved with the emission of neutrons, gamma rays, and 
light charged particles, which carry the information of the dynamical evolution
of the underlying fission dynamics.  
Fission also acts as a source of production for many rare isotopes, 
which can be harvested for use in fundamental research as well as applied areas 
like nuclear medicine and technology. The neutron-rich fission fragment nuclei  
that are produced in a fission process, display a remarkable range of
exotic nuclear phenomena which are in general, investigated experimentally 
through in beam gamma-ray spectroscopic method.

Many scientific investigations have already been 
carried out on the fragment nuclei that are produced in 
$^{252}$Cf~\cite{hamilton1995,hamilton1,hamilton2,zhu1,biswas1,biswas2005} and 
$^{248}$Cm spontaneous fissions (SF) \cite{urban2001,urban2,bentaleb,korgul1}, neutron induced fission \cite{bail2011, gupta2017, mukhopadhyay2012}, 
as well as multiple heavy- and light-ion induced fission reactions\cite{fotiades2002,pantelica2005,wu2004,danu,banerjee}. 
However, it is worth noting that the fission products  
are not the same in each fission reaction. Rather, each of them
is produced in a certain percentage (characteristics of their fission ``yield'')
of the total number of fissions.
As an example, the fission yields of the neutron-rich fragments in the mass A$\approx$100
region is significantly larger in the $^{235}$U($n_{th}$,$f$) fission reaction than those 
from the $^{252}$Cf and $^{248}$Cm SF reactions \cite{ndc}.

It is worthwhile mentioning that the isotopic yield distributions of a fission 
process carry valuable information about that particular fission process, 
and provide the necessary observables for making comparison with the predicted 
results from theoretical models of fission.
Thus, it is of great importance to basic research as well as applications in reactor physics.
For example, the knowledge of these yield distributions provides the key 
data to understand issues related to nuclear fuel waste.
Fission yield data are also of importance in the various nuclear energy 
applications, such as, reactivity or decay heat in nuclear power plants, 
post-irradiation scenarios, neutron-flux determination, and so on.
 
There are several available techniques to estimate experimentally 
the relative isotopic yield distribution of fission fragments. 
In the case of $^{235}$U($n_{th}$,$f$) fission reaction, such studies have already
been carried out using recoil mass separators, time-of-flight measurements using MWPCs (Multi Wire
Proportional Counter) and 
radiochemical analysis \cite{biswas2018, lang1980, schillebeeckx1994, strittmatter1978, rudstam1990}. 
It is to be pointed out here that the majority of the previous investigations were aimed for
extracting the yields of the isotopes belonging to the lighter fragment region. 
The necessary attempt for extracting the relative isotopic yields for the
correlated fission fragments for $^{235}$U($n_{th}$,$f$) through the prompt $\gamma$ ray spectroscopic method 
has not been undertaken prior to the present investigation.
The only available isotopic yield data for this fissioning system based on gamma
spectroscopy is from the work of Mukhopadhyay {\it et al}. \cite{mukhopadhyay2012}. 
The results obtained from this previous investigation were found to be very
limited due to the use of a very small array comprising of only 
two Compton suppressed clover Ge detectors.
In the present investigation, an attempt has been made to extract precisely the 
relative isotopic yield distribution of all the possible even-even fission fragment
pairs populated within the sensitivity limit of the measurement. A large array
of gamma detectors were used for the present investigation and both    
the $\gamma-\gamma$ and $\gamma-\gamma-\gamma$ coincidence data were used
for extracting the yields. The advantage of triple gamma coincidence measurements is that
they can provide sufficient resolving power to analyse the complex gamma radiation following
fission. Identification of new transitions in a selected nucleus can be achieved by gating a known
transitions in this nucleus and/or in the complementary fission fragment.

It is notable here that the prompt $\gamma$-ray spectroscopic method has 
got an added advantage over the other methods for measuring the isotopic yields. 
This is due to the fact
that one can ideally measure the fragments with a unit mass resolution 
through the $\gamma$ ray spectroscopic method. 
However, special care is to be taken for extracting unambiguous yields 
through the spectroscopic method. The factors that make this approach highly sensitive 
are (a) the lack of reasonably developed level schemes in the literature
for certain neutron-rich fragment nuclei, (b) unavoidable yield contributions 
from the precursors’ beta decay, (c) large electron conversion probability 
for certain $\gamma$-transitions, and (d) the presence of low spin milli-second
and micro-second isomers {\it etc.}. In the present work, the uncertainties due 
to all the above mentioned factors have been thoroughly considered and 
the necessary corrections are taken into account within the limit of practical applicability.

The unambiguous results obtained from the present investigation have been
compared with the results from a new type of calculations carried out under the novel 
formalism of isospin conservation (ISCF).
The results from isospin conservation formalism, together with the semi-empirical GEneral description of Fission (GEF)
model calculations \cite{schmidt2016}, have been found to be in fair 
agreement with the experimental findings.

\section{Experimental Setup}
The experiment was performed at the PF1B neutron beam line of the high-flux 
reactor facility at the Institut Laue-Langevin (ILL), Grenoble, France. 
The fission fragments were produced following the induced fission of $^{235}$U
by thermal neutrons and identified through standard $\gamma$ ray spectroscopic
method. The collimated neutron beam flux was 
$\approx$10$^8$~n.~s$^{-1}$~cm$^{-2}$ at the target position. 
The proper collimation of the neutron beam was made using a series of lithium and 
boron collimators mounted upstream from the target position.   
This collimated neutron beam was allowed to impinge 
upon a UO$_2$ target of thickness $\approx$600~$\mu$g/cm$^2$.
The target was about 99.7$\%$ enriched in $^{235}$U.  
The target was sandwiched between thick backings in order to stop the fission 
fragment nuclei and avoid the Doppler shifts of the $\gamma$ peaks.
The gamma rays emitted from the various fission fragments were detected
by the EXILL array ~\cite{exill2017}. The array was comprised of eight EXOGAM 
Clover detectors, six large coaxial detectors from GASP, and the two clovers from the ILL.
The pictorial view of the experimental setup is shown in Fig.~\ref{fig:illsetup}.
The BGO anti-Compton shields were used as Compton suppressors for the EXOGAM 
and GASP detectors. The EXOGAM clovers were mounted in a 90$^{\circ}$ ring around 
the target position. The other detectors were placed in two other rings  
positioned at 45$^{\circ}$ and 135$^{\circ}$.  The total photo-peak efficiency 
for the array was estimated as about 6$\%$ \cite{france2015}.
The data were collected with a trigger-less, digital data acquisition system 
based on 14 bit 100 MHz CAEN digitizers. Details about the electronics and data 
acquisition systems can be found in Ref.~\cite{mutti2013}.
A total of about 2$\times$10$^{10}$ $\gamma-\gamma$ and about 
3.5$\times$10$^{9}$ $\gamma-\gamma-\gamma$ coincidence events were 
recorded during the experiment.
\begin{figure}[h!]
\includegraphics[trim=0.0cm 0.0cm 0.0cm 0.0cm, clip=true, scale=0.6,angle = 0]{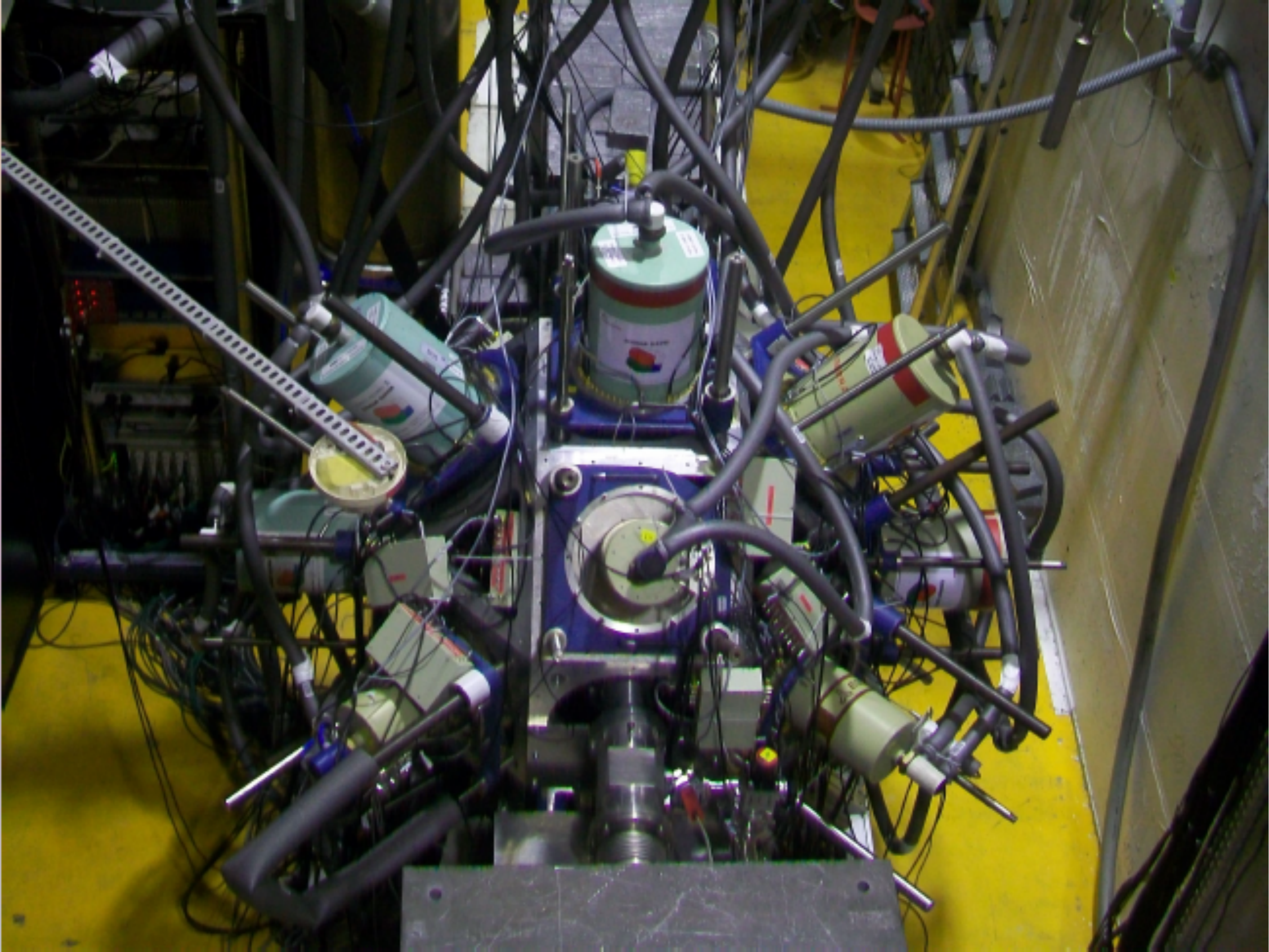}
\caption{
\label{fig:illsetup}
(Color online) A top-view of the experimental EXILL setup at the PF1B neutron beam line area of the high-flux reactor facility at the Institut Laue-Langevin (ILL), Grenoble, France.
}
\end{figure} 
 
\section{Data Analysis and Experimental Results}

The raw data files collected during the experiment were converted into 
structured ROOT tree files for an appropriate prompt coincidence time window 
of 200 $n$s. An in-house data sorting code was developed to
read the ROOT tree files, apply quadratic calibration, generate
spectra for each detector segment as well as for clovers in add-back mode.
The spectra were generated in RADWARE~\cite{rad2} format.
Subsequently $\gamma-\gamma$ coincidence matrices, and $\gamma-\gamma-\gamma$ cubes 
were constructed for further off-line analysis. 
The detailed off-line coincidence spectral analysis was performed using the 
standard analysis software packages, RADWARE \cite{rad2,radford1995} and Tv \cite{theu1993}.
In the present work, the relative isotopic yield distributions of several 
complementary even-$Z$, even-$N$ fission fragment pairs have been measured. 
The complementary fragment pairs that have been analyzed from the present set 
of data are (Mo-Sn), (Zr-Te), (Sr-Xe), (Kr-Ba), and (Se-Ce); thereby
covering a mass range from 80 to 150.
Based on the selection of gating and observed transitions, 
the yields of the complementary fragment pairs 
have been extracted by implementing four different methods of analysis.
The execution of this exercise was needed to gather complete knowledge about
the relative merits and demerits of each of the analysis method, which in turn was 
found to be very helpful in standardizing the best analysis procedure for extracting
the relative isotopic yields of the complementary fission fragment pairs. 
The four different analysis methods have been described one by one in
the follow up discussion.
(a) Method-1:  
In this method, the yield of a particular fragment nucleus has been 
extracted from the area of the peak corresponding to its 
2$_{1}^{+}$ $\rightarrow$ 0$_{1}^{+}$ transition seen in the coincidence
spectrum generated by setting the gate on the 4$_{1}^{+}$ $\rightarrow$ 2$_{1}^{+}$ 
transition of the same nucleus.  \\
(b) Method-2:
In this particular method, the coincidence gates are set on the 
2$_{1}^{+}$ $\rightarrow$ 0$_{1}^{+}$ transitions of the two correlated fission 
fragment nuclei. The coincidence peak yields of the respective 
4$_{1}^{+}$ $\rightarrow$ 2$_{1}^{+}$ transitions of all the possible 
complementary fragments are then extracted simultaneously.\\
(c) Method-3:
The peak yield of the 2$_{1}^{+}$ $\rightarrow$ 0$_{1}^{+}$ transition of the 
fission fragment has been measured from the coincidence spectrum
generated by setting the coincidence gates on 4$_{1}^{+}$ $\rightarrow$ 2$_{1}^{+}$ 
and 2$_{1}^{+}$ $\rightarrow$ 0$_{1}^{+}$ transitions belonging to 
all the other possible complementary fragments.\\
(d) Method-4: 
Any two of the 2$_{1}^{+}$ $\rightarrow$ 0$_{1}^{+}$, 4$_{1}^{+}$ $\rightarrow$ 2$_{1}^{+}$,  and 
6$_{1}^{+}$ $\rightarrow$ 4$_{1}^{+}$ transitions of a fragment nucleus have simultaneously been gated, and the
resultant coincidence spectrum has been generated.
The area of the peak corresponding to the third
transition of the concerned nucleus has been fitted from the coincidence spectrum and 
the yield of the fragment has been extracted. \\
\ \\
It is to be noted here that Method-1 deals with two-fold $\gamma-\gamma$ coincidence 
analysis; whereas the rest of the three methods are based on the analysis
using three-fold $\gamma-\gamma-\gamma$ coincidence data.
As an example, these methods have been applied to the (Zr-Te) complementary fragment pair. The partial level
scheme showing the yrast-line transitions of $^{100}$Zr and $^{134}$Te are shown in Fig.~\ref{fig:levelscheme}.
The representative coincidence spectra obtained under different analysis methods  
is shown in Fig.~\ref{fig:spectrumquality}.
The figure demonstrates that there is gradual improvement in 
the peak-to-total ratio for the concerned peaks
obtained from Method-4, Method-1, Method-3, and Method-2, respectively.
Here, the peak-to-total ratio parameter can be defined as the ratio between the area of a peak and the total area in a given spectrum.
\begin{figure}[h!]
\includegraphics[trim=0.0cm 13.50cm 0.0cm 2.0cm, clip=true, scale=0.6,angle = 0]{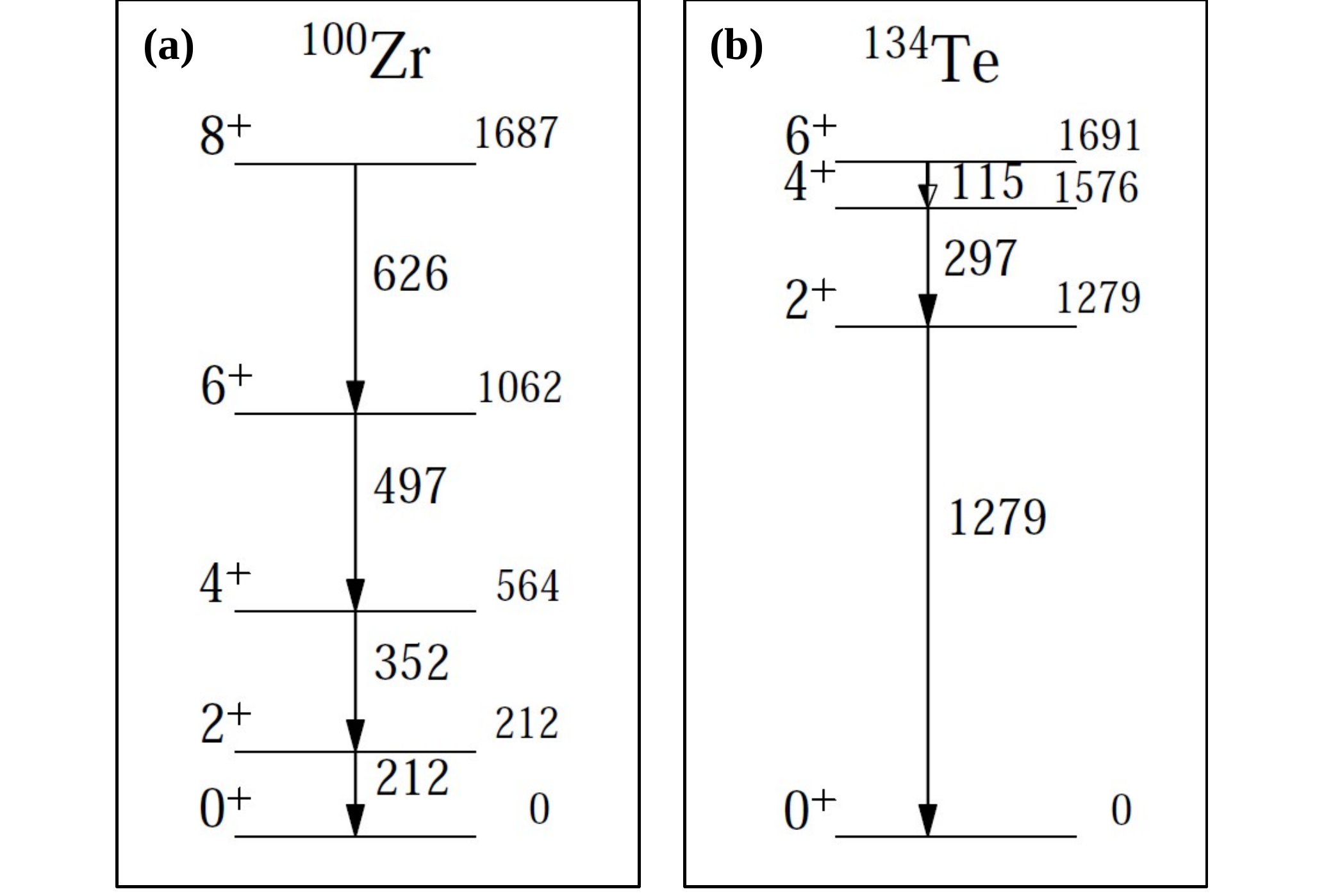}
\caption{
\label{fig:levelscheme}
(Color online) Partial level scheme of (a) $^{100}$Zr and (b) $^{134}$Te, respectively. All the relevant spectroscopic information have been taken from the ENSDF database \cite{nndc}.
}
\end{figure} 

\begin{figure}[h!]
\includegraphics[trim=0.0cm 0.0cm 0.0cm 0.0cm, clip=true, scale=0.6,angle = 270]{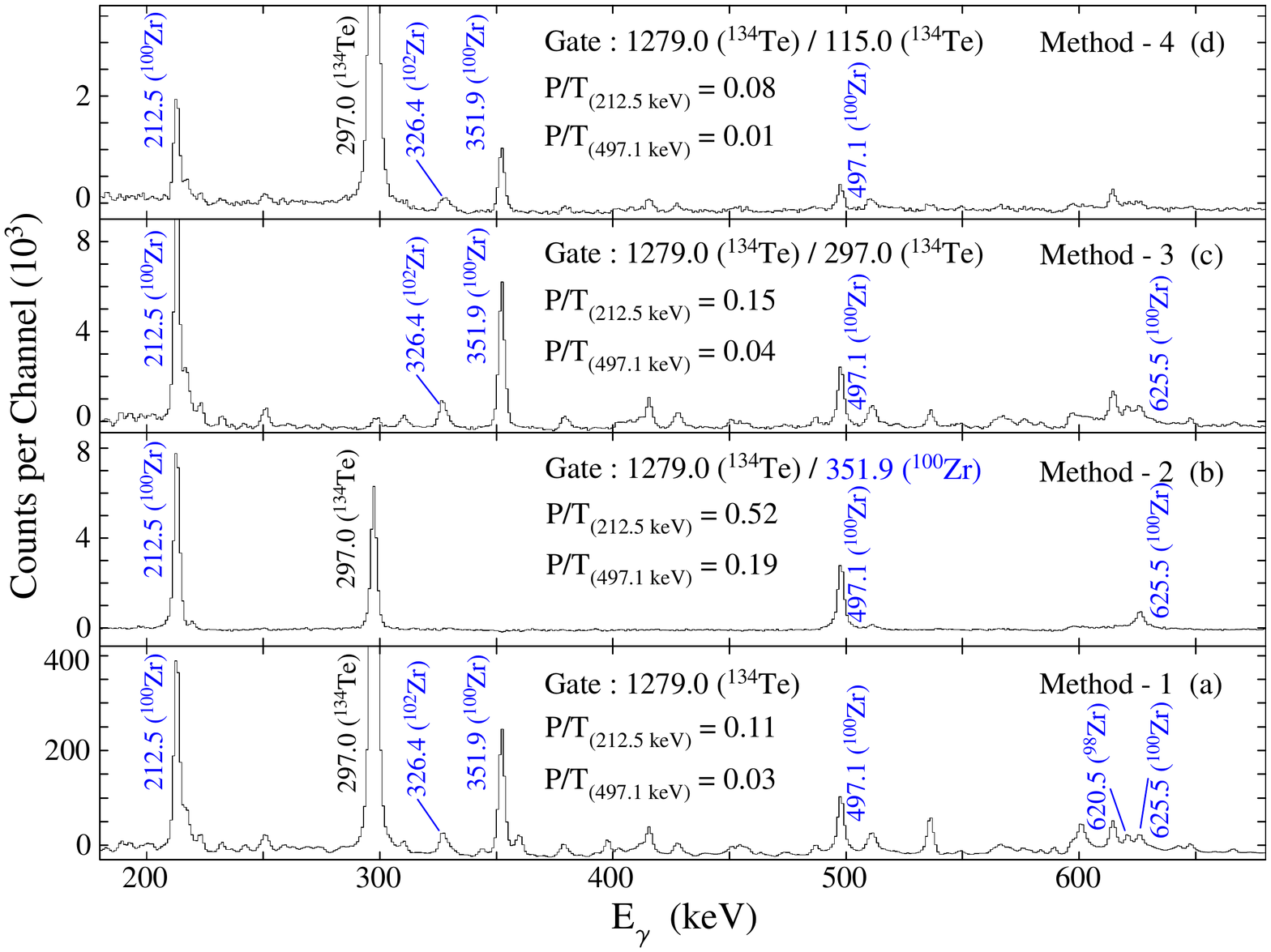}
\caption{
\label{fig:spectrumquality}
(Color online) Representative coincidence spectra obtained by Method-1, Method-2, 
Method-3 and Method-4 are shown in panel (a), (b), (c) and (d) respectively.
For panel (a), the gate is set on 1279.0(2$_{1}^{+}$ $\rightarrow$ 0$_{1}^{+}$)-keV
transition of $^{134}$Te.
For panel (b), the gates are set on 1279.0(2$_{1}^{+}$ $\rightarrow$ 0$_{1}^{+}$)-keV 
transition of $^{134}$Te and 351.9 (4$_{1}^{+}$ $\rightarrow$ 2$_{1}^{+}$)-keV 
transition of $^{100}$Zr. 
For panel (c), the gates are set on 1279.0(2$_{1}^{+}$ $\rightarrow$ 0$_{1}^{+}$)-
and 297.0(4$_{1}^{+}$ $\rightarrow$ 2$_{1}^{+}$)-keV transition of $^{134}$Te.
For panel (d), the gates are set on 1279.0(2$_{1}^{+}$ $\rightarrow$ 0$_{1}^{+}$)-
and 115.0(6$_{1}^{+}$ $\rightarrow$ 4$_{1}^{+}$)-keV transition of $^{134}$Te.
}
\end{figure} 
In order to obtain unambiguous results for the isotopic yields of 
the complementary fission fragment pairs, a detailed analysis of the coincidence data
has been carried out using the aforesaid four different methods. This exercise
has helped us to gather knowledge about the relative merits and demerits of
each of the analysis method which in turn was found to be very helpful in
standardizing the best analysis procedure and obtaining comprehensive
results. A detailed description on the merits and demerits for each of the above mentioned
methods have been put forward in the follow up discussion.

Measurements of the relative isotopic yield distribution of fission 
fragments using prompt gamma ray spectroscopy have been carried out in most of the previous investigations following Method-1, which involves the analysis using two-fold $\gamma-\gamma$ coincidence data.  As mentioned in the aforesaid section, the present investigation 
has been carried out using a large array of gamma detectors. 
Hence, the off-line processing of data into three-fold $\gamma-\gamma-\gamma$ coincidences 
with sufficient statistics could be possible. Such analysis processes ensure better 
selection and identification of the fragments so that the accurate extraction of 
the relative yields of the fission fragments could be possible. 
As noted above, apart from Method-1, all the other methods (Method-2, 3, and 4) 
are based on $\gamma-\gamma-\gamma$ cube analysis. 
In the case of Method-4, there appears to be a serious limitation for extraction 
of the population yield of a fragment nucleus. This is due to the fact that
in the case of a vibrational nucleus, the population strength for the states 
above the 4$^{+}_{1}$ level gets fragmented among the close-lying 
multi-phonon vibrational states. Hence, the strength of 
the 6$^{+}_{1} \rightarrow 4^{+}_{1}$ transition does not simply reflect 
total yield at that particular excitation regime. 
Hence, this method looks to be only applicable in the case of fragments 
having rotational character. The fission fragments having both the low-lying 
rotational and vibrational structures are populated well in the present 
investigation. Hence, Method-4 does not seem to be the appropriate method 
for extracting the isotopic yields. Hence, the results obtained with this method
are not reported in the present work.

In a recent study for the spontaneous fission of 
$^{252}$Cf \cite{musangu2020}, Method-3 has been utilized for the whole set of analysis, and 
for the first time a second mode of fission has been reported. 
But since in this method both the gamma coincidence gates are set 
on one particular fragment, the observed coincidence population statistics
for the complementary fragment nucleus is relatively lower in nature. 
This approach offers a very good scope of selectivity 
for the complementary fragments, but suffers from problems 
in extracting the yields for the probable isotopes having 
poor population strengths. In the present data set, the yields of 
the isotopes such as $^{82}$Se, $^{104}$Zr, $^{98,100}$Mo, $^{134}$Xe, 
and $^{144}$Ce could not be measured due to the aforesaid limitation associated
with Method-3. Method-2 involves the cross gating gamma transitions for selecting
simultaneously both the complementary fragments. Hence, this method not only deals
with the proper identification of the complementary fragments, but also provides 
better enhancement in statistics for the coincidence peaks corresponding 
to the concerned complementary fragments. 

\subsection{Additional corrections associated with the determination of fragment yields}
In the present work, we have incorporated three different types
of corrections for extracting unambiguously the relative isotopic yields of the 
complementary fission fragment pairs. These include the necessary corrections
associated with (a) conversion electron process, (b) contribution from precursors’
beta decay, and (c) presence of milli or micro second isomers. 
It is to be pointed out here that while dealing with Method-2 and -3, the correction
factor for contribution from precursors’ beta decay is not needed as both 
the complementary fission fragments are simultaneously observed in one step
coincidence yield measurement. For the estimation of conversion electron probability, 
the BrIcc code \cite{kibedi2008} has been utilized for all the $\gamma$-transitions having 
energies below 500 keV. It is to be pointed out here that for proper estimation
of the correction factor (required only for Method-1 and -4)
associated with the contribution from the precursors’ beta decay to the 
lower mass isotopes, the corresponding yields from Method-2 have been utilized. 
A comparison between the yield of a fragment nucleus measured from self-gating
and that from the complementary fragment gating (all the even channels are taken into account)
gives the excess contribution from the beta feeding. These correction factors are 
then utilized in both Method-1 and -4. This approach of correction has been made
possible due to the availability of high statistics $\gamma-\gamma-\gamma$ coincidence
data. It is notable here that huge contributions (under Method-1 and -4) 
from the precursors’ beta decay have been observed in the cases of $^{100}$Mo, 
$^{130}$Te, $^{134}$Xe, and $^{144}$Ce. The additional corrections to the yields 
were required due to the presence of low-spin $\mu$-second isomers 
in several fragment nuclei populated in the present work.  
These include the isomeric states belonging to $^{126-132}$Sn, $^{132}$Te,
and $^{136}$Xe. The necessary correction factor due to the presence 
of an isomer is calculated by estimating at first the yield of the particular 
fragment above this concerned isomeric state. 
Following this, the proper estimation for the proportionate decay 
based on the coincidence time window and the isomeric level life-time
has been made. On comparing the coincidence yields of 
the 2$^{+}_{1} \rightarrow 0^{+}_{1}$ transition extracted from 
the gates setting on the transitions lying above and below 
the particular isomeric state, the necessary correction factor is 
estimated subsequently. Apart from the above mentioned corrections, 
additional corrections due to the associated side-feedings have also 
been appropriately incorporated in the present work for 
all the four analysis procedures. The concerned side-feedings are 
due to those transitions which directly decay to the 2$^{+}_{1}$ 
level of a particular fission fragment. It is a matter of common 
fact that nuclei which have low-lying vibrational level structures
are associated with comparably larger amount of side feeding contributions. 
In the present work, it has been observed that the transitions decaying 
from 2$^{+}_{2}$, 3$^{-}_{1}$, and 4$^{+}_{2}$ levels are the dominant 
side-feeding contributors. The estimated side-feedings (in percentages) 
for different fission fragment nuclei are shown in Table~\ref{table:feeding}. 
While dealing with the extraction of isotopic yields of 
the complementary fission fragments, a significant amount of direct feeding 
of the gamma rays from the excited 2$^{+}_{2}$ levels to the ground states 
of the respective fragments have also been observed in a few cases. 
For $^{88}$Se, $^{90}$Kr, $^{96}$Sr, and $^{98}$Zr, the estimated feedings 
from such transitions are found to be (in percentage) 18, 16, 7, and 12, respectively. 
The direct feedings of $E3$ transitions from the 3$^{-}_{1}$ level to the ground 
states have also been observed in a few cases. 
These include the fragments $^{96}$Zr, $^{132}$Sn, and $^{146}$Ba. 
The concerned additional yields of these fragments due to $E3$ feedings   
have been extracted from the known intensity ratios between the 
3$^{-}_{1} \rightarrow 0^{+}_{1}$ and 3$^{-}_{1} \rightarrow 2^{+}_{1}$ transitions
available from the ENSDF database \cite{nndc}.

\subsection{Observations and error analysis}
For the majority of the fission fragment nuclei, it has been observed that the 2$_{1}^{+}$ $\rightarrow$ 0$_{1}^{+}$ 
and 4$_{1}^{+}$ $\rightarrow$ 2$_{1}^{+}$ transitions are nearly of equal intensity 
which suggests that the process of de-excitation occurs predominantly through this cascade.
However, the major limitations for extracting the yields following gamma ray spectroscopic
method arises from the non-availability of detailed level schemes for many of the neutron-rich nuclei. The limitations obviously seem to be more pronounced while dealing with 
$\gamma-\gamma-\gamma$ coincidence analysis for the the non-yrast excited states.
However, one can assume that contributions from any of the unobserved feeding
transitions from the non-yrast states to the ground and 2$^{+}_{1}$ states would
be very feeble. Hence the contributions from such unobserved feedings 
to the final yields of the fission fragments would be negligible.

\begin{figure}[h!]
\includegraphics[trim=0.0cm 0.0cm 0.0cm 0.0cm, clip=true, scale=0.6,angle = 270]{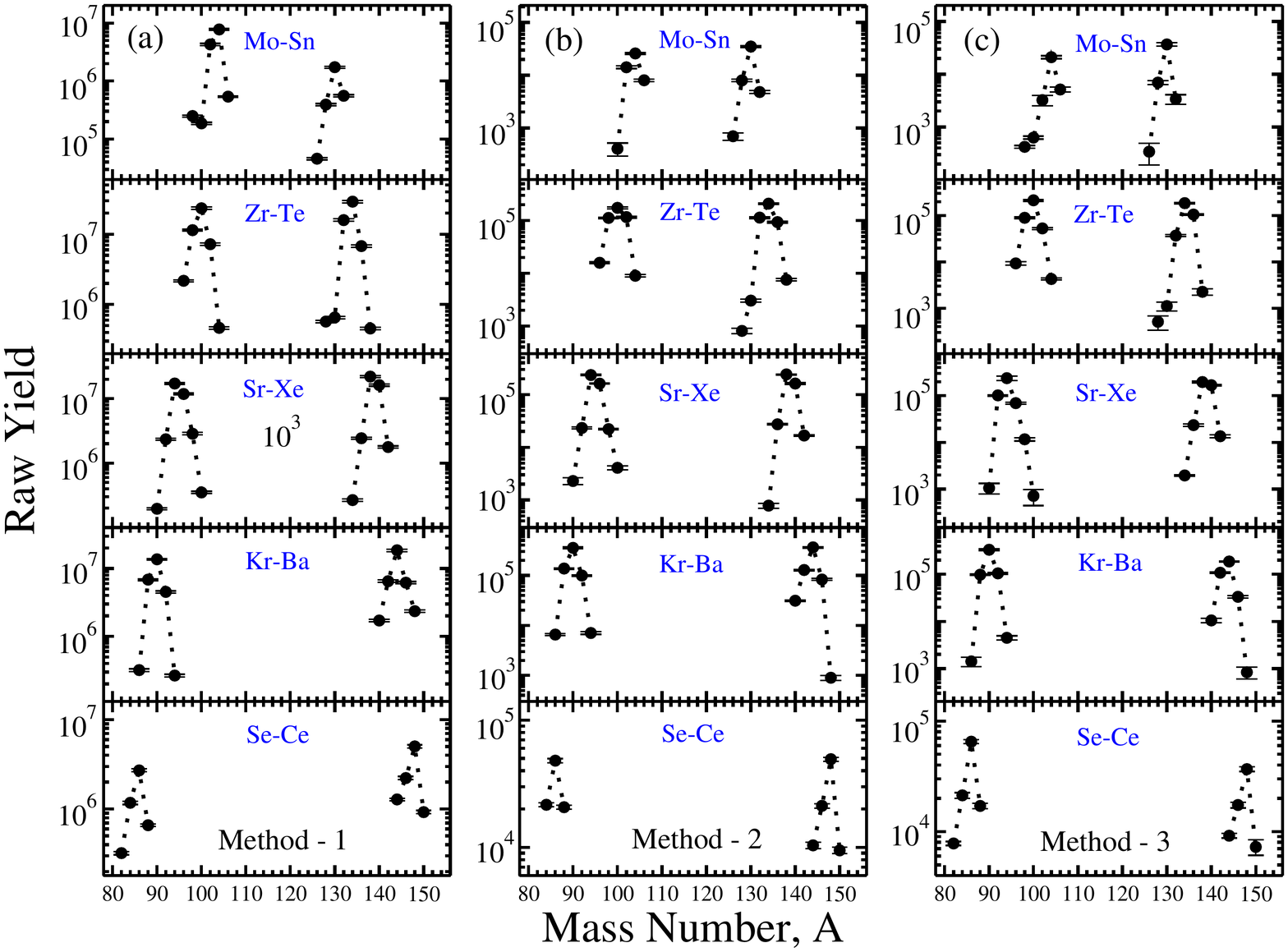}
\caption{
\label{fig:rawyield}
(Color online) Raw isotopic yield distribution of the complementary 
even-even fission fragment pairs extracted from $\gamma-\gamma$ coincidence analysis 
through (a) Method-1, (b) Method-2, and (c) Method-3. See text for details.}
\end{figure} 

\begin{figure}[h!]
\includegraphics[trim=0.0cm 0.0cm 0.0cm 0.0cm, clip=true, scale=0.6,angle = 270]{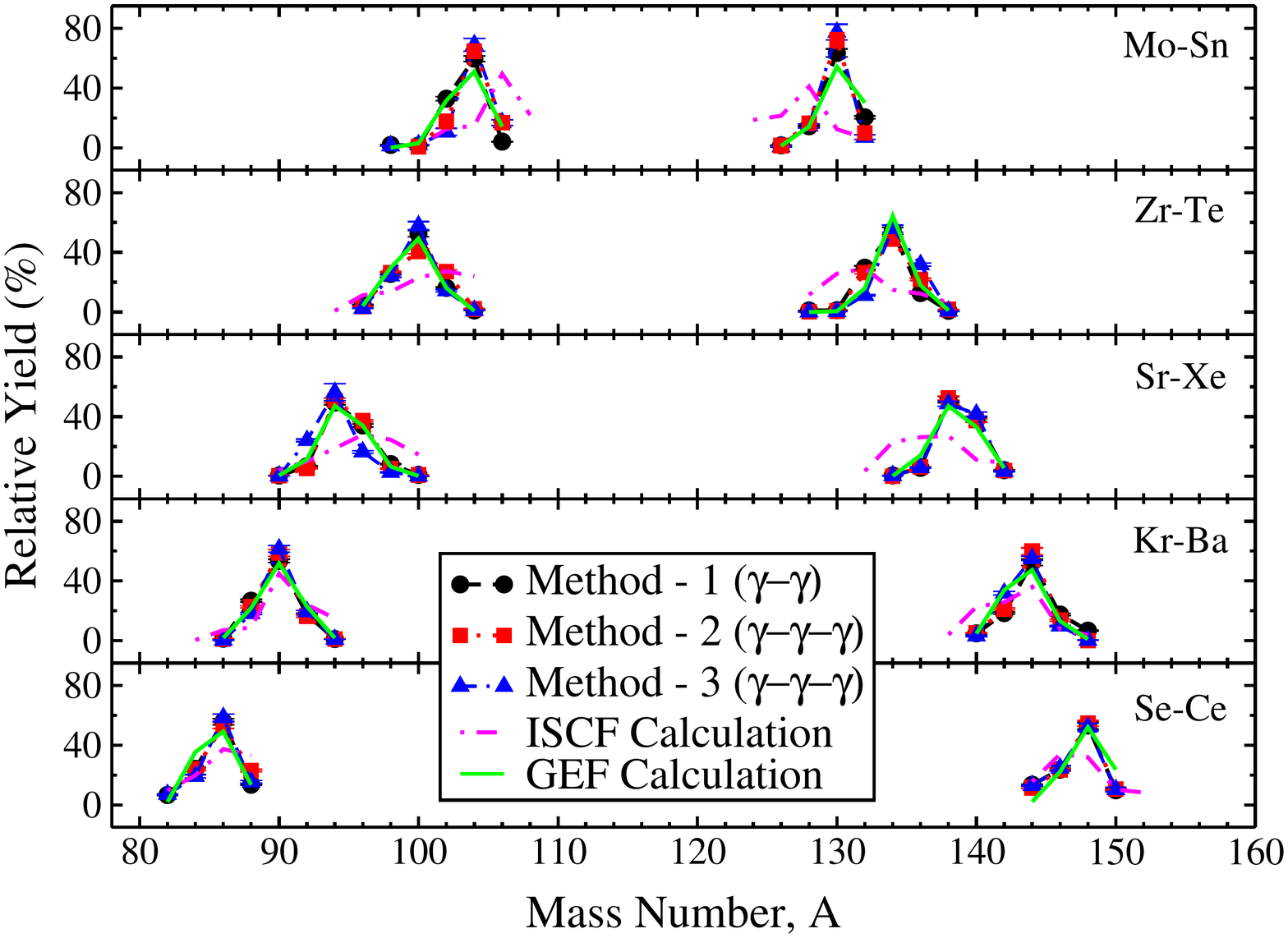}
\caption{
\label{fig:ISCF_GEF}
(Color online) Comparison of relative yield distribution for the complementary 
even-even fission fragment pairs extracted from the analysis 
of Method-1, Method-2, and Method-3. The predicted results from GEF and ISCF calculations
are also shown.
The sum of the relative yields of all the isotopes of a particular fragment has been normalized to 100.
}
\end{figure} 

\begin{figure}[h!]
\includegraphics[trim=0.0cm 0.0cm 0.0cm 0.0cm, clip=true, scale=0.6,angle = 270]{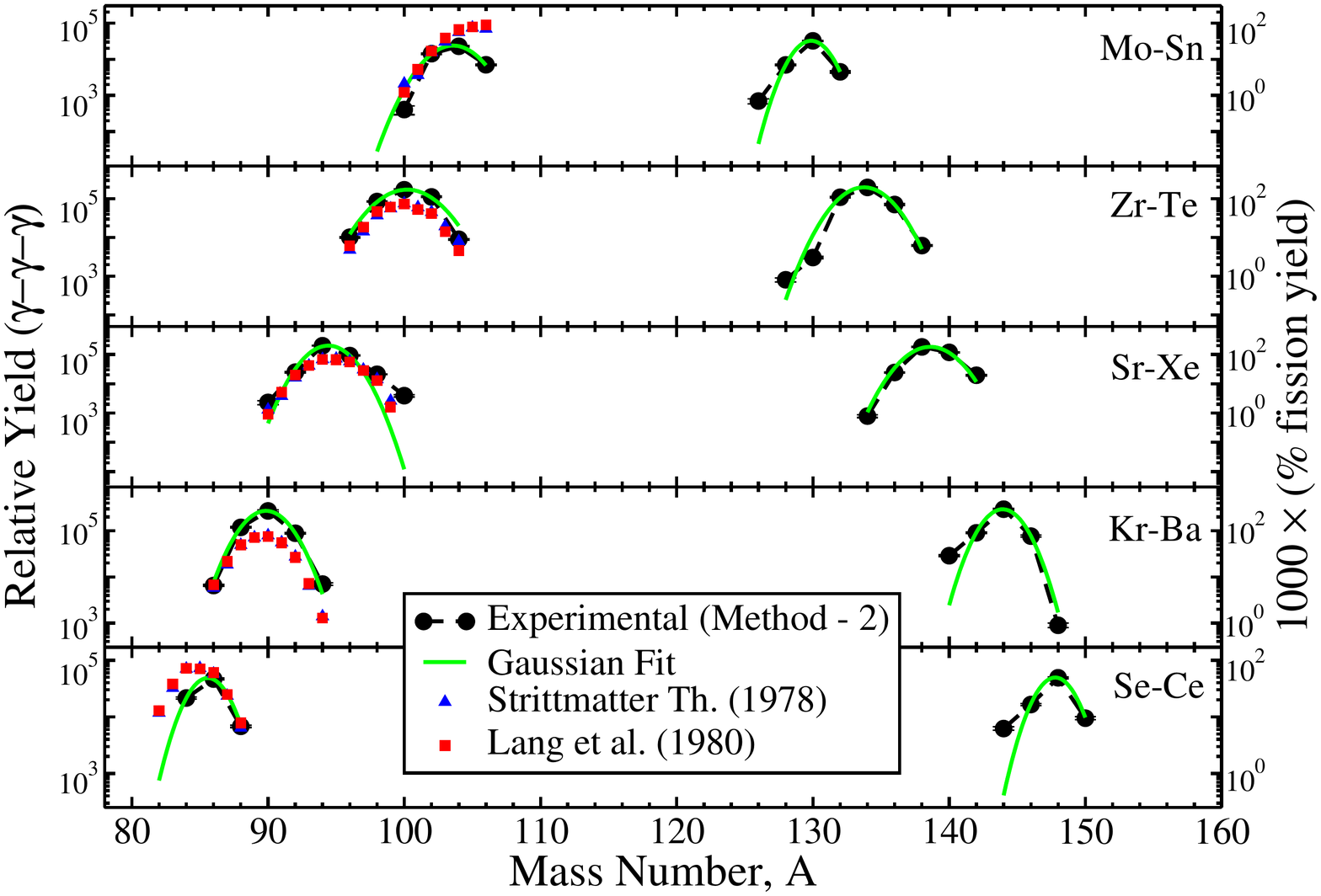}
\caption{
\label{fig:gaussianfit}
(Color online) Comparison of the relative isotopic yield distribution of 
the complementary fission fragment nuclei obtained from the present 
investigation (by using the analysis procedure of Method-2) with those
from the previous investigations. The Gaussian fit to the data has also
been shown. Note the difference in the scale for the left and right ordinates.
The scale of the left ordinate has been used for representing the relative yields 
extracted from the present gamma ray spectroscopic analysis. The right ordinate
scale has been used for representing the yield data obtained from the previous 
measurements using recoil mass separators \cite{lang1980,strittmatter1978}.   
See text for details.
}
\end{figure} 

The raw yield distributions of the complementary fission fragment pairs
obtained from Method-1, Method-2, and Method-3 are shown respectively 
in Figs.~\ref{fig:rawyield} (a), (b), and (c). It is interesting to note that 
the yields for the Sn isotopes extracted from Method-1 are found to be significantly less
than that of the complementary fragment pairs of Mo isotopes. 
The extracted results for the Mo-Sn complementary fragments from Method-2 do not indicate such an obvious difference 
in the yields while Method-3 shows an opposite behavior (i.e. yield of Sn isotopes are more than Mo isotopes).
For the pair of Se-Ce also, this behavior of reversing the trend in yield has also been observed.
These observations brings out the
limitations in the analysis procedure of Method-1 as described in the aforesaid section.    
Similar type of anomalies in the measured yields for the pair of Mo-Sn isotopes has also
previously been reported in the spectroscopic work of $^{238}$U($n$,$f$) reaction 
\cite{wilson2017}, where the data were analyzed using Method-1. 
However, another group verified from direct fission fragment yield measurements that there was no significant yield difference for the Mo-Sn pair in $^{238}$U($n$,$f$) reaction \cite{Ramos2019}.
The observed anomaly in the present and previous spectroscopic measurements
is related to the $\gamma$-multiplicity difference among the correlated fragment pairs of Mo-Sn.
The results from the theoretical calculations of Ref.\cite{Pasca2018} also highlights this observation.
In the present work, the uncertainties (about 3$\%$ has been considered)
due to efficiency corrections of the $\gamma$-transitions have been added in quadrature
to the statistical uncertainties to obtain the final uncertainties associated
with the extracted yields of the fragments. The uncertainties due to the implementation
of various corrections are also included in quoting the final uncertainties associated
with the measured yields of the fragments.
The relative yields (in $\%$) 
of the complementary fragments obtained from Method-1, -2, and -3 are compared
in Fig.~\ref{fig:ISCF_GEF}. The figure indicates marginal difference in yields
of the complementary fragment pairs for all the three methods.
However, the raw data depicted in the Fig.~\ref{fig:rawyield} indicates that there are about 63\%, 5\% and 27\% differences in the
maximum yields of the complementary fragment pair of Mo-Sn corresponding to Method-1, 2, and 3, respectively.
These percentage values have been calculated using the ratio between the difference in the maximum 
peak yields of the pair of complementary fragments and their corresponding mean values.

The comparison of the extracted yields of the pairs of complementary 
fragments obtained from the present (following Method-2) and previous investigations
is shown in Fig.~\ref{fig:gaussianfit}. The previous two investigations were 
carried out by Lang {\it et al.} \cite{lang1980} and Strittmatter {\it et al.} \cite{strittmatter1978} using the Lohengrin mass separator and HIAWATHA recoil 
mass spectrometer, respectively. For making a qualitative comparison between the extracted
yields from the present measurement and the absolute yields from the previous measurements, the absolute 
yield data have been multiplied by a factor of 1000 and plotted accordingly in Fig.~\ref{fig:gaussianfit}.
As can be seen from the figure,
the results from the previous investigations are available for the
lighter fragments of each of the complementary pairs and the patterns of yield distribution appear to collate well
with the results from the present investigation.
The present work thus provides a complete landscape for the relative isotopic
yield distribution pattern of each of the even-even fragment pairs, 
populated within the limit of measurable sensitivity of the present experimental set up, 
following $^{235}$U($n_{th}$,$f$) reaction. Thus, the experimental results
from the present investigation seem to provide a better testing ground for comparing the
results from the models developed under different formalism. The predicted results
from different models have been compared with the results obtained from the present 
investigation in the follow up section.     
The yield data obtained from the present investigation has further been 
fitted using a double Gaussian function. The fitted yield curves
of the different fragment pairs are also shown in Fig.~\ref{fig:gaussianfit}.        
The yield data of a particular pair of complementary fragments has been
fitted with the constraints that the area under both the Gaussian curves
are kept fix while all the other parameters are kept free. 
The Gaussian parameters extracted from the fittings are
enlisted in  Table ~\ref{table:gaussianfit}. 
A mean value of dispersion parameter, $\sigma_{Z}$ = 1.38 
has been obtained from the tabulated data. 
The values of the dispersion parameters obtained from the present work 
are found to be consistent with those obtained from the empirically 
developed A'$_{p}$ model \cite{wahl1988}.

\section{Discussions}
It is generally believed that the isospin remains a 
good quantum number in lighter nuclei having a few protons. 
With increasing proton numbers, the purity of isospin
breaks down due to the emergence of strong Coulomb interactions.
However, following theoretical calculations by Lane and Soper ~\cite{lane1962}, it is highly interesting to note that the purity of isospin gets restored in heavy neutron-rich 
nuclei due to the presence of excess number of neutrons which would dilute the isospin 
impurity of the $N$ = $Z$ core of the heavy nucleus.
Hence, the neutron rich nuclei produced through
spontaneous fissions, heavy-ion induced fusion-fission and other induced-fission reactions
should provide the necessary ground for testing the purity of isospin.
Jain {\it et al.}~\cite{jain2014} proposed the aforesaid idea, and developed the necessary
formalism based on Kelson's arguments~\cite{kelson1969}, for assigning the specific
isospin to neutron-rich fragment nuclei produced in a fission process.
Subsequent calculations for the relative yields of the fission fragments were carried out
based on the assigned isospin and related algebra.
The formalism has already been tested for the relative isotopic yields of
the correlated even-even fission fragments produced through heavy-ion 
induced fusion-fission reactions and thermal neutron induced fission of
$^{245}$Cm ~\cite{garg2017,garg2018,garg2019}. A very good agreement was obtained
between the theoretical and the experimental relative yields for the correlated 
even-even fission fragment pairs. 
In the present work, the isospin conservation formalism has been adopted for new
calculations on the even-even fission-fragment yields from the $^{235}$U($n_{th}$,$f$)
reaction, and the results have been compared with those extracted from 
the high-statistics experimental data.

The values of the total isospin (T) and the third component
of isospin (T$_{3}$) of all the even-even fragment nuclei that are possible 
in the $^{235}$U($n_{th}$,$f$) reaction
have been uniquely assigned (following the prescription proposed by Jain 
and collaborators \cite{garg2017,garg2018,garg2019}), and 
are tabulated in Table ~\ref{table:isospin}.
A pictorial representation of the assigned values of T and T$_{3}$ for different
fragment nuclei is shown in Fig.~\ref{fig:isospinassignment}. 
The assignment of the values is based on two conjectures 
proposed by Kelson~\cite{kelson1969}. These are:
(a) the higher the number of neutrons ($\nu$) emitted in a fission process,
the greater is the probability for the formation of highly excited states 
with T $>$ T$_{3}$, and (b) the fission fragment nuclei are preferably
formed in isobaric analog states (IAS). 
It has already been explained that for the IAS, 
it is sufficient to consider three member isobaric multiplets
(see Table ~\ref{table:isospin}) for assigning a proper isospin value
(T) to the fragment nuclei of a particular mass number~\cite{garg2018}.

\begin{figure}[h!]
\includegraphics[trim=0.0cm 0.0cm 0.0cm 0.0cm, clip=true, scale=0.6,angle = 0]{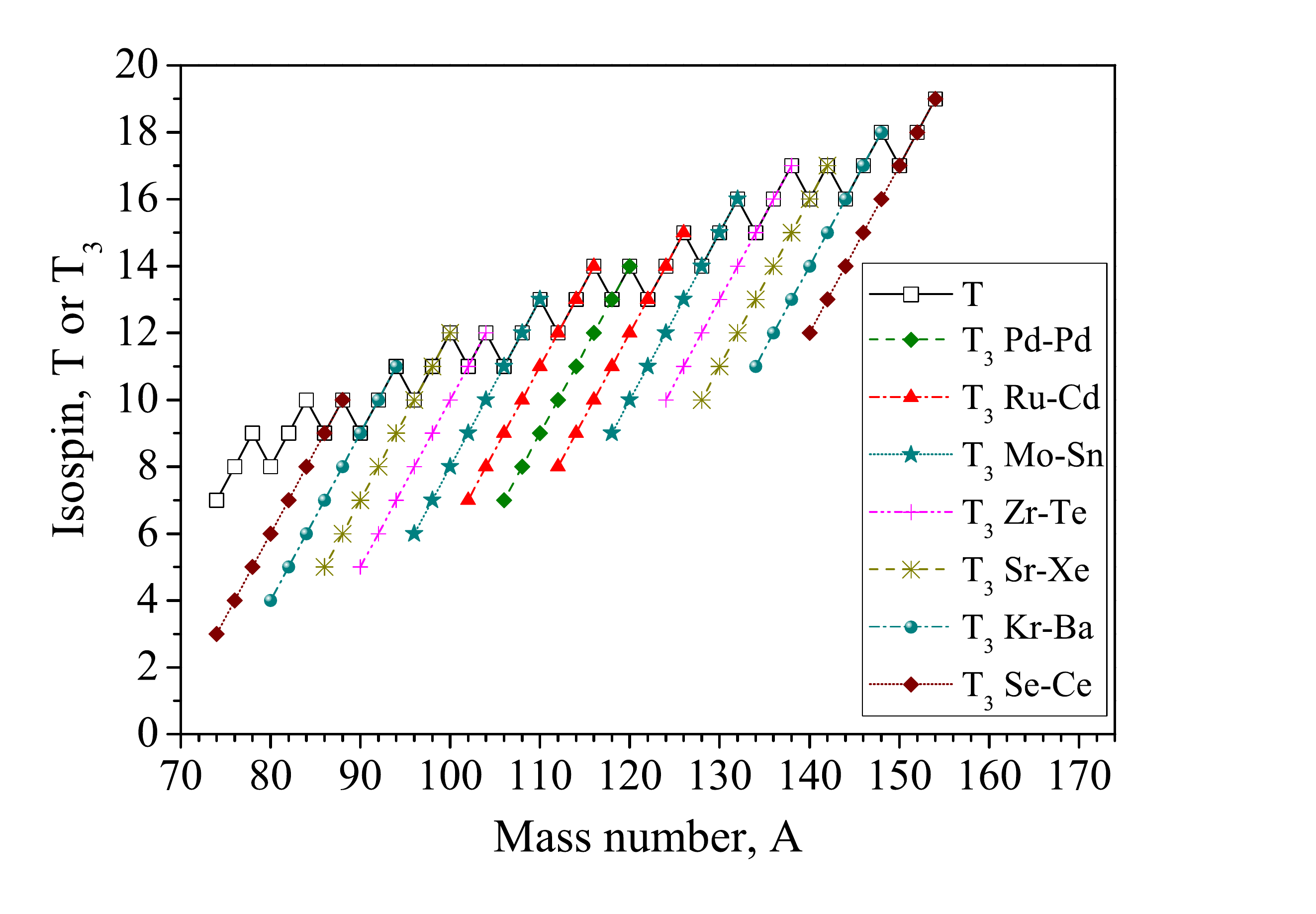}
\caption{
\label{fig:isospinassignment}
(Color online) Depiction of isospin (T or T$_3$) versus mass number (A) 
for the fragment nuclei produced in the $^{235}$U($n_{th}$,$f$) reaction 
after neutron emissions. The open squares on the zig-zag line represent 
the isospin values (T) that have been assigned to each mass number. 
The other symbols represent the T$_3$ values for other fragment masses.
}
\end{figure}

Considering only the isospin part of the total nuclear 
wave function, the yield of a particular pair of even-even fragment nuclei
(represented as F$_{1}$ and F$_{2}$) produced
via ``$\nu$'' number of neutron emission channel is given as \cite{garg2019}:

\begin{equation}
\label{eq:equation}
I_{\nu}  = \left<T_{F_{1}}T_{F_{2}}T_{{3F}_{1}}T_{{3F}_{2}} | T_{RCN} T_{3RCN}\right>^{2} = (CGC)^{2}
\end{equation}

where, $RCN$ denotes the residual compound nucleus that 
has been formed after the emission of $\nu$ number of neutrons from the 
initial compound nucleus ($CN$) such that, $RCN$~=~$F_{1}$+$F_{2}$,
with $T_{RCN}$ $\geq$ $T_{3RCN}$; the isospin and third component of isospin
of $RCN$ is represented by $T_{RCN}$ and $T_{3RCN}$, respectively; 
the terms $T_{F_{1}}$ ($T_{{3F}_{1}}$)
and $T_{F_{2}}$ ($T_{{3F}_{2}}$) denote the isospins (third components of isospins)
for the fragments, $F_{1}$ and $F_{2}$, respectively; 
the term on the right hand side of the above equation
represents the square of the Clebsch Gordon Coefficients~(CGC).

The calculated isotopic yields for the even-even complementary 
fission fragment nuclei are depicted along with the experimental results 
in Fig.~\ref{fig:ISCF_GEF}. It is obvious from the figure that the calculated 
yields are in reasonable agreement with those extracted from the experimental data. 
This leads us to conclude that the isospin remains an approximately
good quantum number for the neutron-rich fission fragments, 
produced in thermal neutron induced fission of $^{235}$U.
It is worthwhile mentioning that the calculation using the isospin conservation formalism
is valid for the compound nucleus (CN) based fissioning systems only. 
Further, the formalism assumes that all the neutrons from the compound nucleus are emitted in one single step, thereby making no distinction between the pre- and post-scission neutrons.

Further calculations were carried out under the framework of 
the semi-empirical GEF model\cite{schmidt2016}.
The GEF model is based on a general approach to nuclear fission and provides
a considerable insight into the physics of the fission process. Using the GEF model, 
most of the fission observables can be calculated with a precision 
that complies with the needs for applications in nuclear technology. 
The simulation code based on the GEF model is designed as a Monte-Carlo formalism
that follows all the quantities of the fissioning systems with their correlations
and dependencies. 
It is evident from Fig.~\ref{fig:ISCF_GEF} that the GEF model reproduces the 
experimentally deduced relative isotopic yield distributions of the correlated 
even-even fragment nuclei quite well.

Thus it can be summarized that the overall agreement 
between the calculated results from the semi-empirical based GEF model 
calculations and the experimental data is quite evident. 
Moreover, it should be noted that the results from the standard GEF model 
calculations are found to be in reasonable agreement with those obtained from the new calculations 
based on isospin conservation formalism for the $^{235}$U($n_{th}$,$f$) reaction.

\section{Conclusion}

Correlated neutron rich fission-fragment nuclei 
produced in the $^{235}$U($n_{th}$,$f$) reaction have been investigated 
following prompt high-resolution $\gamma$ ray spectroscopy
method. The high-statistics coincidence data were obtained during the EXILL campaign
at ILL, Grenoble, France. The relative isotopic yield distributions of the even-even fission 
fragments have been extracted combining the results from $\gamma-\gamma$ and 
$\gamma-\gamma-\gamma$ coincidence analysis.
The extraction of yields of the correlated fragment pairs have been carried out
using four different methods of analysis. Merits
and demerits of the methods have been discussed in detail. The best suitable method of analysis
for the extraction of unambiguous yields of the correlated fragment pairs
has been identified.
An attempt has been made for interpreting the
experimental results by a theoretical model calculation based on isospin conservation
formalism (ISCF). However, only a reasonable agreement has been observed between the
theoretical and experimental results presented in this paper, suggesting an improvement in the
model is required to explain the fission yield distribution.
The experimental results have also been compared with those obtained
from the semi-empirical GEF model calculations.
The predicted results from the two different theoretical approaches
are found to be in reasonable agreement with each other.

\section{Acknowledgments} 
The authors are grateful to the EXOGAM collaboration for
the loan of the detectors, to GANIL for providing assistance during
installation and dismantling, and to the INFN Legnaro laboratory
for the loan of the GASP detectors.
Two of the authors (A.D. and A.C.) gratefully
acknowledge the financial assistance received from the DAE-BRNS, Govt. of India 
(Project Sanction No.: 37(3)/14/17/2016-BRNS) for carrying out this work.
One of the authors (A.K.J) acknowledges the support from Science and Engineering Research Board (Govt. of India) of a Grant No. CRG/2020/000770.
One of the authors (S. Garg) acknowledges the partial support
by the National Natural Science Foundation of China (Grant No. U1932206)
and by the National Key Program for S $\&$ T Research and Development (Grant No. 2016YFA0400501).

\begin{table}
\caption{
The extracted Gaussian parameters obtained from the fitting of
the relative isotopic yield distributions of the complementary
fission fragment pairs of $^{235}$U($n_{th}$,$f$).
}
\label{table:gaussianfit}
\begin{ruledtabular}
\begin{tabular}{cccccc}
Sl. No	& Fission Fragment	& Z	& $\sigma$$_{Z}$	& A$_{P}$	& N$_{P}$/Z	\\
\hline
1	& Se			& 34	& 1.24			& 85.58		& 1.49		\\
2	& Kr			& 36	& 1.43			& 89.84		& 1.49		\\
3	& Sr			& 38	& 1.26			& 94.42		& 1.49		\\
4	& Zr			& 40	& 1.82			& 100.21	& 1.49		\\
5	& Mo			& 42	& 1.52			& 103.59	& 1.44		\\
6	& Sn			& 50	& 1.07			& 129.87	& 1.59		\\
7	& Te			& 52	& 1.56			& 133.74	& 1.56		\\
8	& Xe			& 54	& 1.44			& 138.61	& 1.55		\\
9	& Ba			& 56	& 1.27			& 143.92	& 1.55		\\
10	& Ce			& 58	& 1.22			& 147.79	& 1.54		\\
\end{tabular}
\end{ruledtabular}
\end{table}

\begin{longtable}{cccccc}
\caption{\label{table:feeding}
List of fission fragment nuclei having measurable amount of feeding 
from the different excited levels, 
other than that from the 4$^{+}_{1}$, to the 2$^{+}_{1}$ level.
The nuclei those have the feeding component from the 4$^{+}_{1}$ level
only are not enlisted in the table. The measured feeding contributions (in $\%$) from
different components are shown in the rightmost column of the table.  
All the other relevant spectroscopic quantities in the table 
has been taken from the ENSDF database \cite{nndc}.
}
\endfirsthead
\caption[]{continued...} \\
\hline
Nuclei	& Ground state 	& J$_i$$^\pi$ $\rightarrow$ J$_f$$^\pi$	& Feeding	& J$_i$$^\pi$ $\rightarrow$ J$_f$$^\pi$	& Feeding	\\
	& transitions	&					& transitions	&					& $\%$		\\
	& (keV)		&					& (keV)		&					&		\\
\hline
\endhead
\endfoot
\hline
\endlastfoot
\hline
\hline
Nuclei	& Ground state 	& J$_i$$^\pi$ $\rightarrow$ J$_f$$^\pi$	& Feeding	& J$_i$$^\pi$ $\rightarrow$ J$_f$$^\pi$	& Feeding	\\
	& transitions	&					& transitions	&					& $\%$		\\
	& (keV)		&					& (keV)		&					&		\\
\hline
$^{86}$Se	& 704.3	& 2$_{1}^{+}$ $\rightarrow$ 0$_{1}^{+}$	& 863.4		& 4$_{1}^{+}$ $\rightarrow$ 2$_{1}^{+}$	& 90		\\
		&	&					& 694.6		& 2$_{2}^{+}$ $\rightarrow$ 2$_{1}^{+}$	& 10		\\
$^{88}$Se	& 589.4	& 2$_{1}^{+}$ $\rightarrow$ 0$_{1}^{+}$	& 961.9		& 4$_{1}^{+}$ $\rightarrow$ 2$_{1}^{+}$	& 83		\\
		&	&					& 653.5		& 2$_{2}^{+}$ $\rightarrow$ 2$_{1}^{+}$	& 17		\\
$^{88}$Kr	& 775.2	& 2$_{1}^{+}$ $\rightarrow$ 0$_{1}^{+}$	& 868.4		& 4$_{1}^{+}$ $\rightarrow$ 2$_{1}^{+}$	& 95		\\
		&	&					& 802.1		& 2$_{2}^{+}$ $\rightarrow$ 2$_{1}^{+}$	& 2		\\
		&	&					& 1328.9	& 4$_{2}^{+}$ $\rightarrow$ 2$_{1}^{+}$	& 3		\\
$^{90}$Kr	& 707.0	& 2$_{1}^{+}$ $\rightarrow$ 0$_{1}^{+}$	& 1123.4	& 4$_{1}^{+}$ $\rightarrow$ 2$_{1}^{+}$	& 43		\\
		&	&					& 1057.1	& $\rightarrow$ 2$_{1}^{+}$		& 23		\\
		&	&					& 655.5		& 2$_{2}^{+}$ $\rightarrow$ 2$_{1}^{+}$	& 34		\\
$^{92}$Kr	& 768.8	& 2$_{1}^{+}$ $\rightarrow$ 0$_{1}^{+}$	& 1035.3	& 4$_{1}^{+}$ $\rightarrow$ 2$_{1}^{+}$	& 72		\\
		&	&					& 1296.7	& 4$_{2}^{+}$ $\rightarrow$ 2$_{1}^{+}$	& 28		\\
$^{92}$Sr	& 814.9	& 2$_{1}^{+}$ $\rightarrow$ 0$_{1}^{+}$	& 858.4		& 4$_{1}^{+}$ $\rightarrow$ 2$_{1}^{+}$	& 81		\\
		&	&					& 1370.0	& 3$_{1}^{-}$ $\rightarrow$ 2$_{1}^{+}$	& 19		\\
$^{94}$Sr	& 836.9	& 2$_{1}^{+}$ $\rightarrow$ 0$_{1}^{+}$	& 1309.1	& 4$_{1}^{+}$ $\rightarrow$ 2$_{1}^{+}$	& 52		\\
		&	&					& 1089.4	& 3$_{1}^{-}$ $\rightarrow$ 2$_{1}^{+}$	& 43		\\
		&	&					& 1812.7	& 4$_{2}^{+}$ $\rightarrow$ 2$_{1}^{+}$	& 5		\\
$^{96}$Sr	& 815.0	& 2$_{1}^{+}$ $\rightarrow$ 0$_{1}^{+}$	& 977.8		& 4$_{1}^{+}$ $\rightarrow$ 2$_{1}^{+}$	& 64		\\
		&	&					& 1160.6	& 4$_{2}^{+}$ $\rightarrow$ 2$_{1}^{+}$	& 11		\\
		&	&					& 692.0		& 2$_{2}^{+}$ $\rightarrow$ 2$_{1}^{+}$	& 19		\\
		&	&					& 1037.3	& 3$_{1}$ $\rightarrow$ 2$_{1}^{+}$	& 6		\\
$^{96}$Zr	& 1750.4& 2$_{1}^{+}$ $\rightarrow$ 0$_{1}^{+}$	& 1106.8	& 4$_{1}^{+}$ $\rightarrow$ 2$_{1}^{+}$	& 32		\\
		&	&					& 146.6		& 3$_{1}^{-}$ $\rightarrow$ 2$_{1}^{+}$	& 68		\\
$^{98}$Zr	& 1222.9& 2$_{1}^{+}$ $\rightarrow$ 0$_{1}^{+}$& 620.5		& 4$_{1}^{+}$ $\rightarrow$ 2$_{1}^{+}$	& 66		\\
		&	&					& 583.2		& 3$_{1}^{-}$ $\rightarrow$ 2$_{1}^{+}$	& 27		\\
		&	&					& 1053.8	& 4$_{3}^{+}$ $\rightarrow$ 2$_{1}^{+}$	& 4		\\
		&	&					& 824.5		& 4$_{2}^{+}$ $\rightarrow$ 2$_{1}^{+}$	& 2		\\
		&	&					& 521.6		& 2$_{3}^{+}$ $\rightarrow$ 2$_{1}^{+}$	& 1		\\
$^{100}$Zr	& 212.5	& 2$_{1}^{+}$ $\rightarrow$ 0$_{1}^{+}$	& 351.9		& 4$_{1}^{+}$ $\rightarrow$ 2$_{1}^{+}$	& 97		\\
		&	&					& 665.9		& 2$_{2}^{+}$ $\rightarrow$ 2$_{1}^{+}$	& 2		\\
		&	&					& 1185.4	& $\rightarrow$ 2$_{1}^{+}$		& 1		\\
$^{102}$Zr	& 151.7	& 2$_{1}^{+}$ $\rightarrow$ 0$_{1}^{+}$	& 326.4		& 4$_{1}^{+}$ $\rightarrow$ 2$_{1}^{+}$	& 90		\\
		&	&					& 1090.8	& 3$_{1}^{+}$ $\rightarrow$ 2$_{1}^{+}$	& 10		\\
$^{104}$Mo	& 192.2	& 2$_{1}^{+}$ $\rightarrow$ 0$_{1}^{+}$	& 368.4		& 4$_{1}^{+}$ $\rightarrow$ 2$_{1}^{+}$	& 84		\\
		&	&					& 620.2		& 2$_{2}^{+}$ $\rightarrow$ 2$_{1}^{+}$	& 4		\\
		&	&					& 836.3		& 3$_{1}^{+}$ $\rightarrow$ 2$_{1}^{+}$	& 12		\\
$^{106}$Mo	& 171.5	& 2$_{1}^{+}$ $\rightarrow$ 0$_{1}^{+}$	& 350.6		& 4$_{1}^{+}$ $\rightarrow$ 2$_{1}^{+}$	& 81		\\
		&	&					& 538.8		& 2$_{2}^{+}$ $\rightarrow$ 2$_{1}^{+}$	& 8		\\
		&	&					& 713.6		& 3$_{1}^{+}$ $\rightarrow$ 2$_{1}^{+}$	& 11		\\
$^{130}$Sn$^{a}$& 1221.2& 2$_{1}^{+}$ $\rightarrow$ 0$_{1}^{+}$	& 774.3		& 4$_{1}^{+}$ $\rightarrow$ 2$_{1}^{+}$	& 91		\\
		&	&					& 807.0		& 2$_{2}^{+}$ $\rightarrow$ 2$_{1}^{+}$	& 9		\\
$^{132}$Sn$^{a}$& 4041.1& 2$_{1}^{+}$ $\rightarrow$ 0$_{1}^{+}$	& 375.1		& 4$_{1}^{+}$ $\rightarrow$ 2$_{1}^{+}$	& 98		\\
		&	&					& 310.7		& 3$_{1}^{-}$ $\rightarrow$ 2$_{1}^{+}$	& 2		\\
$^{132}$Te$^{a}$& 974.0	& 2$_{1}^{+}$ $\rightarrow$ 0$_{1}^{+}$	& 697.0		& 4$_{1}^{+}$ $\rightarrow$ 2$_{1}^{+}$	& 98		\\
		&	&					& 690.9		& 2$_{2}^{+}$ $\rightarrow$ 2$_{1}^{+}$	& 2		\\
$^{136}$Xe$^{a}$& 1313.0& 2$_{1}^{+}$ $\rightarrow$ 0$_{1}^{+}$	& 381.3		& 4$_{1}^{+}$ $\rightarrow$ 2$_{1}^{+}$	& 89		\\
		&	&					& 976.5		& 2$_{2}^{+}$ $\rightarrow$ 2$_{1}^{+}$	& 9		\\
		&	&					& 812.6		& $\rightarrow$ 2$_{1}^{+}$		& 2		\\
$^{138}$Xe	& 588.8	& 2$_{1}^{+}$ $\rightarrow$ 0$_{1}^{+}$	& 483.7		& 4$_{1}^{+}$ $\rightarrow$ 2$_{1}^{+}$	& 98		\\
		&	&					& 875.2		& 2$_{2}^{+}$ $\rightarrow$ 2$_{1}^{+}$	& 2		\\
$^{140}$Xe	& 376.6	& 2$_{1}^{+}$ $\rightarrow$ 0$_{1}^{+}$	& 457.6		& 4$_{1}^{+}$ $\rightarrow$ 2$_{1}^{+}$	& 92		\\
		&	&					& 927.9		& 3$_{1}^{+}$ $\rightarrow$ 2$_{1}^{+}$	& 5		\\
		&	&					& 1136.7	& 3$_{1}^{-}$ $\rightarrow$ 2$_{1}^{+}$	& 3		\\
$^{144}$Ba  & 199.3	& 2$_{1}^{+}$ $\rightarrow$ 0$_{1}^{+}$	& 330.8		& 4$_{1}^{+}$ $\rightarrow$ 2$_{1}^{+}$	& 97		\\
		&	&					& 638.9		& 3$_{1}^{-}$ $\rightarrow$ 2$_{1}^{+}$	& 3		\\
$^{146}$Ba  & 181.0	& 2$_{1}^{+}$ $\rightarrow$ 0$_{1}^{+}$	& 332.4		& 4$_{1}^{+}$ $\rightarrow$ 2$_{1}^{+}$	& 91		\\
		&	&					& 640.1		& 3$_{1}^{-}$ $\rightarrow$ 2$_{1}^{+}$	& 9		\\
$^{144}$Ce	& 397.4	& 2$_{1}^{+}$ $\rightarrow$ 0$_{1}^{+}$	& 541.2		& 4$_{1}^{+}$ $\rightarrow$ 2$_{1}^{+}$	& 66		\\
		&	&					& 844.8		& 3$_{1}^{-}$ $\rightarrow$ 2$_{1}^{+}$	& 34		\\
\hline
\end{longtable}
\vskip -0.7 cm
$^{a}$ indicates that isomeric correction has been taken into account.


\begin{longtable*}{cccccccc}
\caption{\label{table:isospin}
Third component of isospin (T$_{3}$) and total 
isospin (T) values of all the possible even-even
fission fragments (FF) produced in the $^{235}$U($n_{th}$,$f$) reaction.
}
\endfirsthead
\caption[]{continued...} \\
\hline
A & Nucleus & T$_{3}$ & Nucleus & T$_{3}$ & Nucleus & T$_{3}$ & T \\
\hline
\endhead
\endfoot
\hline
\endlastfoot
\hline
\hline
A & Nucleus & T$_{3}$ & Nucleus & T$_{3}$ & Nucleus & T$_{3}$ & T \\
\hline
86 & $^{86}$Se &  9 & $^{86}$Kr & 7 & $^{86}$Sr & 5 &  9\\
88 & $^{88}$Se & 10 & $^{88}$Kr & 8 & $^{88}$Sr & 6 & 10\\
90 & $^{90}$Kr &  9 & $^{90}$Sr & 7 & $^{90}$Zr & 5 &  9\\
92 & $^{92}$Kr & 10 & $^{92}$Sr & 8 & $^{92}$Zr & 6 & 10\\
94 & $^{94}$Kr & 11 & $^{94}$Sr & 9 & $^{94}$Zr & 7 & 11\\
96 & $^{96}$Sr & 10 & $^{96}$Zr & 8 & $^{96}$Mo & 6 & 10\\
98 & $^{98}$Sr & 11 & $^{98}$Zr & 9 & $^{98}$Mo & 7 & 11\\
100& $^{100}$Sr& 12 & $^{100}$Zr& 10& $^{100}$Mo& 8 & 12\\
102& $^{102}$Zr& 11 & $^{102}$Mo& 9 & $^{100}$Ru& 7 & 11\\
104& $^{104}$Zr& 12 & $^{104}$Mo& 10& $^{104}$Ru& 8 & 12\\
106& $^{106}$Mo& 11 & $^{106}$Ru& 9 & $^{106}$Pd& 7 & 11\\
108& $^{108}$Mo& 12 & $^{108}$Ru& 10& $^{108}$Pd& 8 & 12\\
110& $^{110}$Mo& 13 & $^{110}$Ru& 11& $^{110}$Pd& 9 & 13\\
112& $^{112}$Ru& 12 & $^{112}$Cd& 8 & $^{112}$Pd& 10& 12\\
114& $^{114}$Ru& 13 & $^{114}$Cd& 9 & $^{114}$Pd& 11& 13\\
116& $^{116}$Ru& 14 & $^{116}$Cd& 10& $^{116}$Pd& 12& 14\\
118& $^{118}$Sn&  9 & $^{118}$Cd& 11& $^{118}$Pd& 13& 13\\
120& $^{120}$Sn& 10 & $^{120}$Cd& 12& $^{120}$Pd& 14& 14\\
122& $^{122}$Te$^{a}$&  9 & $^{122}$Sn& 11& $^{122}$Cd& 13& 13\\
124& $^{124}$Te& 10 & $^{124}$Sn& 12& $^{124}$Cd& 14& 14\\
126& $^{126}$Te& 11 & $^{126}$Sn& 13& $^{126}$Cd& 15& 15\\
128& $^{128}$Xe& 10 & $^{128}$Te& 12& $^{128}$Sn& 14& 14\\
130& $^{130}$Xe& 11 & $^{130}$Te& 13& $^{130}$Sn& 15& 15\\
132& $^{132}$Xe& 12 & $^{132}$Te& 14& $^{132}$Sn& 16& 16\\
134& $^{134}$Ba& 11 & $^{134}$Xe& 13& $^{134}$Te& 15& 15\\
136& $^{136}$Ba& 12 & $^{136}$Xe& 14& $^{136}$Te& 16& 16\\
138& $^{138}$Ba& 13 & $^{138}$Xe& 15& $^{138}$Te& 17& 17\\
140& $^{140}$Ce& 12 & $^{140}$Ba& 14& $^{140}$Xe& 16& 16\\
142& $^{142}$Ce& 13 & $^{142}$Ba& 15& $^{142}$Xe& 17& 17\\
\hline
\hline
\end{longtable*}
\vskip -0.7 cm
$^a$ indicates those nuclei which have been included for the sake of completeness
of the table. However, the parameters corresponding to those nuclei
do not affect the calculations, and have not been plotted in Fig.~\ref{fig:isospinassignment}.


\end{document}